\documentstyle[preprint,12pt,aps,prl]{revtex}
\tightenlines
\begin{document}
\newcommand{\nvo}{NaV$_{2}$O$_{5}$}
\newcommand{\cvo}{CaV$_{2}$O$_{5}$}
\newcommand{\cm}{cm$^{-1}$}
\newcommand{\cgo}{CuGeO$_3$}
\newcommand{\nxvo}{Na$_x$V$_{2}$O$_{5}$}
\title{Low-energy excitations in NaV$_2$O$_5$.}
\author{M.J.Konstantinovi\'c $^{\ast}$, J. C. Irwin}
\address{Simon Fraser University, Physics Department, 8888 University 
drive, Burnaby, B.C. V5A1S6 Canada} 
\author{M.Isobe and Y.Ueda}
\address{ $^b$ Institute for Solid State Physics, The University of Tokio, 
7-22-1 Roppongi, Minato-ku, Tokio 106, Japan} 

\maketitle

\begin{abstract} 
In the (ab) polarized Raman scattering spectra of {\nvo} single crystals, 
measured with 647.1 nm laser line at $T < T_c$, we found two 
modes at 86, and 126 {\cm} not previously reported. These two modes,
together with 66, and 106 {\cm} modes, make an array of four 
low-energy equidistant 
modes below the energy onset of the continuum at
about 132 {\cm}. 
All four modes
are strongly suppressed by increasing Na deficiency, 
indicating their nonvibrational origin and
the existence of a quantum phase transition at critical
Na deficiency between 3 and 4\%. 
These results question current understanding of {\nvo} as 
quasi one-dimensional Heisenberg antiferromagnet.
\end{abstract}


Sodium vanadate, {\nvo}, is a mixed valence compound 
($V^{4+}:V^{5+}=1:1$) with a structure consisting
of vanadium-oxygen ($VO_5$) pyramids, that are connected
via common edges and corners to form layers in the the ab plane. The structure
can be described as an array of parallel ladders (running along the {\bf b}
 axis) that are coupled to form a trellis lattice, see Fig. 1. Each rung is made of 
V-O-V bond, and contains one valence electron
 donated by the sodium atoms which are
situated between layers as intercalants.

This compound
exhibits a phase transition at $T_c=34 K$ \cite {a1} that is
 usually referred to in the
literature as a charge-ordering (CO) phase transition. 
From the mixed-valence point of view its existence is not unusual, since 
a common consequence of the mixed valence is the appearance of 
charge-ordering, which is a class of metal-insulator 
transitions
\cite {a2}.
However, the phases on both sides of the transition in {\nvo} are 
insulators, and a very interesting interplay between charge and spin dynamics
results in the phase-transition discovered in this compound. Indeed, despite 
the fact 
that the vanadium ions have uniform valence of $+4.5$ at room temperature
\cite {a3,a4}
(indicating the quarter-filled structure of V-O-V rungs \cite {a6}, 
and suggesting a metalic state!), the
magnetic susceptibility of {\nvo} is in excellent agreement with Bonner-Fisher
curve for a one-dimensional antiferromagnet (AF) \cite {a1}. 
The one-dimensionality of the magnetic ordering is
realized along the {\bf b} axis which is perpendicular to the 
V-O-V rungs \cite {a1}.
Furthermore, at low temperatures, the susceptibility decreases rapidly to zero 
in accord with a spin-liquid ground state, and suggesting the existence of a 
spin-Peierls (SP) transition. However, since the physical properties 
(e.g. transition temperature) of {\nvo}
are insensitive to an applied magnetic field \cite {a1,a7},
the SP scenario can be ruled out. 

Accordingly, several theoretical investigations of the role of electron correlations 
(intersite Coulomb interactions) in charge dynamics, and/or charge-ordering
have been presented \cite {a81,a82,a83,a84,a85,a86}. 
So far, almost all proposed charge configurations ("in-line", "zigzag", 
or a "ladder-like") effectively lead to an 1D Heisenberg AF 
with a spin-excitation gap, and a spin-liquid ground state.
In the spin-cluster model \cite {a8a}, which is different in that respect,
the spin-gap arises from a finite number of spins of the isolated cluster. 
All concepts were tested by comparison with experimental
data with some success, but no consistent picture has emerged yet. 
In fact, the central issue appears to be the energy scale of the CO
in {\nvo} and how it should manifest itself 
in optical \cite {a91,a92,a94,a95,a96}, and neutron scattering spectra 
\cite {a10,a101}.
This concerns both the origin of optical 
transitions in the 0.8-5 eV energy range, as well as the origin of
the low-frequency electronic excitations (observed in both IR and Raman 
spectra), and the modes associated with a spin dynamics. 

In Raman spectra of {\nvo}, the phase transition
manifests itself through the appearance of various new Raman active
modes \cite{a94,a95,a96} below the phase transition temperature.
The high-temperature phonon excitations in the Raman spectra \cite {a96}
are well understood in terms of the crystal symmetry (space group Pmmn) and
of the selection rules as determined from structural analysis \cite{a4}.
The origin of the newly activated Raman modes is, however,
still unclear mainly due to a lack of understanding
of the low-temperature {\nvo} crystal structure \cite {a12}.
Of particular interest are the low-frequency modes at 66, 106, and
132 {\cm} found in the Raman spectra of the low temperature
phase \cite{a94,a95,a96}.
These modes are of special interest because of their proximity to spin-gap
energy ($\sim 65-85$ {\cm} \cite{a10,a101}), and their potential magnetic origin.
The 66 and 106 {\cm} modes were previously assigned as magnetic bound states
\cite{a95},
mainly based on the slow increase (almost perfectly linear) of their intensities
with decreasing temperature (similar behavior is found for the two-magnon 
bound state at 32 {\cm} in $CuGeO_3$ \cite{a12a} ), and 
because of their insensitivity to the magnetic field.
This assignation is supported by most of the theoretical models that
predict a chain type effective magnetic ordering in {\nvo}. 
However, the energy of the 66 {\cm} mode is very close to the
spin gap. According to the recent neutron scattering study the spin-gap mode
coincides in energy with 66 {\cm} mode at ($Q_a=integer, Q_b=Q^{ZC}$), and 
at ($Q_a=half-integer, Q_b=Q^{AF}$).
If this mode is truly a magnetic
bound state of two spin-gap excitations, it is difficult to understand why
its binding energy is close to (equal to), the spin-gap 
energy itself. Two questions naturally arise: Is the 
one-dimensional AF Heisenberg model relevant for the magnetic dynamics
of {\nvo}, and are those low-frequency Raman-active modes really the
magnetic bound states?
Clearly, further experiments are necessary 
in order clarify present understanding of physical properties 
and Raman spectra of {\nvo}.

Here, we present the temperature dependent
Raman spectra of {\nxvo}, $1\ge x \ge 0.96 $. Four modes are found
in the low-temperature, $T<T_c$, (ab) polarized Raman spectra of {\nvo}
that are separated by a common energy of 20 {\cm}.
Two of them, at 66 and 86 {\cm}, are also
detected in recent neutron scattering experiments \cite {a101}. 
Simultaneous activity
of the modes in the Raman and neutron scattering spectra, 
suggest that, at least, the 
66 {\cm} mode is not the magnetic bound state.
The equal energy separation of the low-frequency modes 
and a strong suppression of their energy and intensity by Na
deficiency, question current
understanding of the low-temperature magnetism in {\nvo} 
as being simple 1D alternating Heisenberg AF.

Polarized Raman experiments were performed on {\nxvo} single crystals (size
$\sim$\ $1\times3\times1$~mm$^3$ along {\bf a}, {\bf b}, and {\bf c})
prepared as described in Refs. \cite{a1,a11}. As excitation source we used
647.1 nm line from Kr$^+$ ion laser.  The beam, with an average power of
5 mW, was focused (spot diameter $\sim80\mu$m) on (001) surfaces of the
crystals.
The spectra were measured in a quasi-backscattering geometry using a
{\sc dilor} triple monochromator equipped with a LN$_2$ cooled CCD camera.

In Fig. 2 we present the typical Raman scattering spectra 
of nominally pure {\nvo} in (aa) and (ab)
polarized configurations, measured at T=6 K with 647.1 nm laser line. 
The modes
in the (aa) spectra are in agreement with previous reported data but are
more intense due to a resonance effect \cite{a13}. However, in (ab) spectra the 
two
new modes at 86 and 126 {\cm} are found. These modes were not previously
observed in the Raman spectra measured with 488 and 514.5 nm laser lines 
\cite{a94,a95,a96}.
Altogether, the modes at 66, 86, 106, and 126 {\cm} make an array of four 
low-energy equidistant 
modes, separated by 20 {\cm}. Besides, all four modes have
energies lower then the onset of the continuum that starts at about
132 {\cm}, which is twice the energy of the lowest mode, see Inset of 
Fig. 2. 
A four modes show strong temperature dependence. In fact, their
intensities strongly reduce at the phase transition temperature,
forming a broad band at higher temperatures, see Fig.3.
Also, while all four modes have comparable intensities
in the (ab) polarized configuration, the 86 and 126 {\cm} modes are strongly 
suppressed
in the (aa) and (bb) polarized geometries with respect to the 66 and 106 
{\cm} modes (Fig. 2).
As we already mentioned, the simultaneous activity of the 66 and 86 {\cm} modes
in both neutron and Raman scattering spectra, make assignation of the
66 {\cm} mode as a magnetic bound state highly questionable,
since singlet-singlet 
transitions cannot be observed with a neutron scattering.
Possible interpretation of this mode as one-magnon excitation \cite{a94} 
seems also unlikely,
since no change of either intensity or energy is found in 
the magnetic fields up to 12 T, which is in contrast to the expected
linear splitting for a singlet-triplet transition.
The next possibility is that 66 {\cm} excitation is a vibrational mode; 
folded phonon for example. 
If so, the equidistant energy separation between modes in (ab) polarization 
would then be
a simple coincidence, which we believe is not the case as discussed
in the next paragraph.

In order to further examine the low-frequency dynamics of {\nvo}
we analyze the effects associated with sodium deficiency.
Sodium deficiency strongly influences the properties of \nvo \cite{a1,a11}.
With increasing sodium deficiency, the CO phase transition temperature
decreases and disappears with Na content around 0.97 (3\%\ deficiency)
\cite{a18}.
Fig. 3 shows the low-frequency Raman spectra of nominally pure {\nvo} as
a function of temperature (upper panel), and several
Na-deficient {\nxvo} samples at T=6 K (lower panel) in the (aa) scattering
geometry. For sodium deficiencies above 3\% the modes associated with
the CO phase completely 
vanish, in a good agreement
with earlier data showing a full suppression of the phase transition at
about this concentration \cite{a18}. 
The modes also shift in energy as the Na concentration is reduced 
by about 6-7\% in clear contrast to other (aa) low-temperature modes 
(shifts are $ \sim 1 \%$). 
In the (ab) scattering
geometry these modes behave in a similar way, except that now the 
86 and 106 {\cm} modes cannot be observed for $[Na] \le 0.99$,
probably because of their relatively small intensity. 
Decreasing $[Na]$ introduces the holes in the V-O planes since it
transforms the $V^{4+}$ (S=1/2) ions into $V^{5+}$ (S=0). 
This effect is similar to nonmagnetic-magnetic substitution of ions.
Similar mode suppression is observed in the Raman spectra of 
Zn-substituted {\cgo}
\cite{a19}, where the singlet mode (the
magnetic bound state) shifts to lower energies with increasing Zn doping, 
demonstrating 
renormalization of the spin-gap (a suppression of the phase transition is observed
by substitution of non-magnetic ions for Cu in the spin-Peierls compound
{\cgo} \cite{a20}).
In this case the effect is caused by the presence of unpaired spins, introduced
by breaking of
the spin-singlets upon substitution.
Thus, although we question here the assignation of 66 and 106 {\cm} modes
as a magnetic bound states, sodium deficient Raman spectra 
indicate that low-frequency modes in {\nvo} are not vibrational modes,
since phonon energies should not be so sensitive to small 
impurity concentrations.

From the neutron 
scattering spectra \cite{a101} it has been argued that the 66 and 86 {\cm}
modes have a common origin (two spin-gap modes from two unequal 
magnetic chains), but on the other hand, the Raman spectra indicate
that 66 and 106 {\cm} make pair.  
Therefore, having all this in mind,
it is possible that all four modes have the same origin 
(note, that the weak structure around
106 {\cm} seems to exist in the neutron scattering spectra \cite{a101}).
If so, this question our present understanding of physical properties of 
{\nvo} since, the equidistant mode separation is not inherent property of 
1D Heisenberg AF model.  
The quasi-1D Heisenberg or spin-phonon model has been predicted to have a
spectrum consisting of multiple soliton-antisoliton boundstates, with the
number of boundstates increasing as the inter-chain coupling gets weaker.
However, such a model does not naturally predict 4 equally spaced states
\cite {a13a}.

Still, the presence 
of a strong singularity at $2 \Delta_s$ in the Raman spectra
is the important indication of the low-dimensional magnetism. 
The features that
represent the onset of a two-magnon continuum at $2 \Delta_s$ have been 
observed in the Raman spectra of {\cgo} \cite {a12a}, and {\cvo} \cite {a16},
in the form of strong asymmetric lines with a tail towards higher energies.
The both compounds are quasy-1D magnetic systems with a spin-gap. 
{\cgo} is a 1D AF with a spin gap which opens due to alternation of
exchange constants along the chain. {\cvo} is a weakly coupled dimmer system,
 i.e a two-leg-ladder
compound with much stronger exchange along the rungs than along the legs.
In {\nvo} a similar structure is found around 132 {\cm} 
which could represent the onset of two-magnon 
continuum (see Inset Fig. 2). If so, the spin-gap in {\nvo} should be 66 
{\cm}, and there is a mode at this energy in both neutron and Raman spectra. 
The spin-gap mode is a singlet-triplet transition 
which should split in an applied magnetic field (which is not observed), 
unless the triplet ($S^z$) degeneracy is already somehow removed?
On the other hand, its activity in (ab) polarized scattering configuration 
indicates that the magnetic order in {\nvo} is not the chain-type like
{\cgo}, where two-magnon excitation is observed only with incident
and scattered electromagnetic fields polarized parallel to the chain direction 
(note that in a ladder case the Fleury and Loudon exchange scattering mechanism 
\cite {a17}, allows the two-magnon Raman scattering in a crossed 
configuration as well).

Finally, let us return to Fig. 3.
The most remarkable effect is the similarity between 
the temperature dependent Raman spectra
of pure {\nvo}, and Na deficient, T=6 K Raman spectra. 
If the behavior of any of these modes can be related to order parameter
of the {\nvo} phase transition, then the order parameter
must have the same scaling for either T or x. 
At the moment, it is not clear what causes such an effect,
but we believe that it indicates the influence of quantum criticality to the
phase transition of {\nvo}. 

In conclusion, we find four low-frequency modes 
separated by a common energy of 20 {\cm},
that can be associated with the CO phase of {\nvo}. 
These modes are strongly suppressed with Na deficiency indicating
their nonvibrational origin and the existence of a quantum phase transition
at critical
Na deficiency between 3 and 4\%. 
The equal energy separation of the low-frequency modes 
and a strong suppression of their energy and intensity with Na
deficiency, question current
description of the low-temperature magnetism in {\nvo} 
as being simple 1D alternating Heisenberg AF.

{\bf Acknowledgments}

We thank I. Affleck for fruitful discussions.
This work is supported by Natural Sciences and Engineering Research
Council of Canada.

$^{\ast} mkonstan@sfu.ca$

\begin{figure}
\caption
{ Schematic representation of the NaV$_2$O$_5$ crystal structure in the
(001) and (010) planes. Effective representation of parallel ladders coupled 
in a trellis lattice is also shown.}
\label{fig1}
\end{figure} 
\begin{figure}
\caption
{The low-frequency (aa) and (ab) polarized Raman scattering spectra of 
NaV$_2$O$_5$ at 6 K,
measured using $\lambda$ =647.1 nm laser line. Inset:
The (ab) Raman spectra in extended frequency range
at temperatures above and below CO phase transition.
}
\label{fig2}
\end{figure} 
\begin{figure}
\caption
{upper panel: The low-frequency (aa) polarized Raman scattering spectra at 
various
temperatures between 10 K and 35 K.
lower panel: The (aa) polarized Raman scattering 
spectra of samples with different Na content, measured
at T=6 K. Inset: The Raman spectra of nominally pure NaV$_2$O$_5$ at various 
temperatures up to 100 K.
} 
\label{fig3}
\end{figure} 

\end{document}